# Dependence of the Switching Current Density on the Junction Sizes in Spin Transfer Torque

Chun-Yeol You[1] and Myung-Hwa Jung[2,a]

[1]Department of Physics, Inha University, Incheon 402-751, Korea

[2]Department of Physics, Sogang University, Seoul 121-742, Korea

We investigate the dependence of switching current density on the junction sizes in the spin transfer torque nanopillar structures by using micromagnetic simulations. While the macro spin model predicts weak dependence of switching current density on the junction sizes, we find that the switching current density is a sensitive function of the junction sizes. It can be explained with the complicated spin configurations and dynamics during the switching process. The detail spin configurations and dynamics are determined by spin wave excitation with the finite wave vector, which is related with the exchange coupling energy and junction shape.

---

[a]Author to whom correspondence should be addressed. E-mail : mhjung@sogang.ac.kr

# I. INTRODUCTION

The switching current density is one of the important quantities in the development of the spin transfer torque magnetic random access memory (STT-MRAM) for the next generation non-volatile memory applications. The switching current density is determined by the instability conditions of the spin wave, which is excited by the STT. In addition, it is widely accepted that the switching current density is determined by the various physical parameters such as spin polarization, saturation magnetization, and shape of magnetic tunneling junction (MTJ), based on the macro spin model [1,2,3]. In the macro spin model, the contribution of the exchange stiffness is usually ignored, because the contribution of exchange energy is zero. However, we have recently found that the switching current density is a sensitive function of the exchange stiffness constant by using the micromagnetic simulations [4]. We have also found that the detailed spin configuration and dynamics are important in the realistic switching processes. The detailed spin configuration and dynamics are determined by the exchange stiffness constant and the shapes of the MTJs, which play an important role in the determination of switching current density.

In this study, we investigate the dependence of switching current density on the junction sizes for various exchange stiffness constants. By using public domain micromagnetic simulator, Object-Oriented MicroMagnetic Framework (OOMMF) [5] with the public STT extension module [6,7], we calculate the switching current density of typical MTJ structure with various lateral junction sizes. We find that the dependence of switching current density on the junction sizes is not a simple monotonic function, but it shows complicated behaviors. The results can be explained with the detailed spin configurations and weakly quantized spin wave with finite wave vector.

## II. SIMULATIONS

We consider the typical STT-MRAM structure with an insulating barrier between exchange-biased synthetic ferrimagnet layer of $F_3$/NM/$F_2$ and free ferromagnetic layer $F_1$ as shown in Fig. 1 [8,9]. The saturation magnetization and the ferromagenet thicknesses of $F_{1,2,3}$ layers are $1.3 \times 10^6$ A/m and 2 nm, respectively. The thicknesses of normal metal (NM) and insulator (I) layers are both 1 nm. The cross section of the nanopillar is an ellipse of $a \times b$ nm$^2$, where the long axis length is $a$ and the short axis length is $b$. In this simulation, we vary $a$ from 30 to 120 nm with $b$ = 20, 30, and 40 nm, and the cell size is $1 \times 1 \times 1$ nm$^3$. The exchange stiffness constants are considered to be $A_{ex}$ = 1.0, 2.0 and $3.0 \times 10^{-11}$ J/m, because the reported experimental data of most interesting materials such as CoFeB are in these ranges, depending on the composition and fabrication details [10,11,12]. For simplicity, no crystalline anisotropy energy is considered and the Gilbert damping constant is fixed to be $\alpha$ = 0.02. The exchange bias field of $4 \times 10^5$ A/m is assigned to the long axis of the ellipse (+$x$-direction) for the $F_3$ layer. We consider only the in-plane STT contribution and ignore the out-of-plane STT contribution. We apply positive current, where electrons flow from free to reference layer, which prefers antiparallel state. The pulse duration time is 10 ns, and we check the switching status after 2 ns. By repeating the procedure, we determine the switching current density for the parallel to antiparallel states. All micromagnetic simulations are done at zero temperature, and thus the thermal excited contributions are ignored in this study. More details of micromagnetic simulations can be found elsewhere [6].

## III. RESULTS AND DISCUSSIONS

The simulation results of the switching current densities $J_c$ for various junction sizes are depicted in Fig. 2 (a)~(c) and Fig. 3 (a)~(c), as a function of the long axis length of the ellipse ($a$ = 30 ~ 120 nm) for different short axis length of the ellipse ($b$ = 20, 30, and 40 nm) with different exchange stiffness constant ($A_{ex}$ = 1.0, 2.0, and 3.0×10$^{-11}$ J/m). Figure 2 shows the $J_c$ variations for different $b$ values with a fixed $A_{ex}$ value at each panel, while Figure 3 shows the $J_c$ variations for different $A_{ex}$ values with a fixed $b$ value at each panel as the same data set. In Fig. 2 (a) with $A_{ex}$ = 1.0×10$^{-11}$ J/m, $J_c$ oscillates strongly for $b$ = 30 nm, oscillates weakly for $b$ = 20 nm, and increases monotonically for $b$ = 40 nm. These patterns are changed for $A_{ex}$ = 2.0 ×10$^{-11}$ J/m in Fig. 2 (b), and they are slowly increased for $A_{ex}$ = 3.0 ×10$^{-11}$ J/m in Fig. 2 (c). Such tendencies are not easy to be explained with simple macro spin model. In order to get better insight, we replot the same data in Figs. 3 (a)~(c) for fixed $b$ values. The $J_c$ variations are quite different for the short axis length of $b$, in spite of the small change of $b$. The variation is most serious for $b$ = 30 nm seen in Fig. 3 (b), and it is somewhat monotonic for $b$ = 40 nm seen in Fig. 3 (c). Such strong dependence of $J_c$ on the junction size and exchange stiffness is our main finding in this study. It should be noted that the switching current density without consideration of junction size and exchange stiffness, which is calculated from macro spin model, is $J_{c0}$ = 1.84×10$^{11}$ A/m$^2$. This value is much smaller than all of our micromagnetic simulation results.

Let us explain the physical reasons of such variation of $J_c$ based on the macro spin model including the contribution of exchange stiffness term with finite spin wave vector $k$ [13,14,15,16,17,18]. The switching current density $J_c$ is given by

$$J_c \sim \frac{\alpha}{a_1}\left[H_{eff} + \frac{1}{2}\left(N_y + N_z - 2N_x\right)M_s + \frac{2A_{ex}}{\mu_0 M_s}k^2\right]. \tag{1}$$

Here, $N_{x,y,z}$ denote the demagnetization factors of the free layer ($N_z \gg N_y \geq N_x$ for thin ellipse), and $H_{eff}$ is the effective field including external field, stray field, Oersted field, and perpendicular STT field-like term. $\alpha$ is the Gilbert damping constant and $a_1 = \eta_p \frac{\hbar}{2e\mu_0 M_s d_s}$. Here, $\eta_p$, $d_s$, $M_s$, $\mu_0$, and $\hbar$ are the spin polarization of the polarizer layer, thickness of the free layer, saturation magnetization, permeability of the vacuum, and reduced Plank's constant, respectively. According to Eq. (1), the contribution of exchange stiffness with spin wave vector $k$ is clear for thin films. The easiest excitation occurs with $k = 0$, which is a uniform mode. However, we calculate with the finite size nanopillar structures, and thus the $k$ value is limited by the junction dimension. Therefore, the $k$ value is weakly quantized to minimize the exchange and demagnetization energies. Such weak quantization can roughly explain the oscillatory behaviors of $J_c$.

Now, we discuss the details of the spin configurations during the switching process in order to support above explanations. Let us focus a $60 \times b$ nm$^2$ ellipse with $A_{ex} = 1.0 \times 10^{-11}$ J/m, which indicates the red dashed circle in Fig. 2 (a). The switching current densities $J_c$ vary from 2.68 to $4.75 \times 10^{11}$ A/m$^2$, when the short axis lengths $b$ are different. The $60 \times 40$ nm$^2$ case shows the lowest $J_c$ value, and the $60 \times 30$ nm$^2$ case shows the highest $J_c$ value. To reveal the reason, we apply $J = 3.0 \times 10^{11}$ A/m$^2$, smaller than $J_c$ for $b = 20$ and 30 nm but larger than $J_c$ for $b = 40$ nm, which points out a blue horizontal arrow in Fig. 2 (a). We depict the time dependence of normalized $x$-component of magnetization $M_x$ at positions A, B, C, and D (see Fig. 1 for the definition of each position), and total $M_x$ in Fig. 4 (a)~(c). For $b = 20$ and 30 nm, the magnetization at positions A and B shows large oscillations for whole time. For example, the amplitude of oscillation at position A for $b = 20$ nm in Fig. 4 (a) changes between -1.0 to 0.8, indicating that the magnetization of position A is almost switched. Then, the amplitude of oscillation decreases at positions B and C. Finally, at position D which is the center of the ellipse, no oscillation is observed in spite of the strong oscillation of the off-center positions. As a result, the total oscillation is finite and the switching is not occurred. This situation is similar to the $b = 30$ nm case in Fig. 4 (b). Even though the applied current density $J$ is smaller than $J_c$ for $b = 20$ and 30 nm, because $J$ is larger than $J_{c0}$ (= $1.18 \times 10^{11}$ A/m$^2$), the observed large oscillations are not surprising. On the other hand, for the $b = 40$ nm case in Fig. 4 (c) the magnetization is switched around 9 ns.

For more details, we perform the fast Fourier transform (FFT) of time dependent magnetization as shown in Figs. 5 (a)~(c). The corresponding frequency dependencies are shown in Figs. 4 (a)~(c). There are several points we should address about the Fourier analysis. First, it is clearly shown that the ellipse with larger $b$ shows more complicated modes, implying that the oscillation is not coherent. This result is reasonable when we consider the junction size. It is more manifested in Fig. 6, that will be discussed later. Second, the main peak frequencies decrease from 22, 20, to 17 GHz for $b = 20$, 30, and 40 nm, respectively. Since all physical parameters are same except the short axis length $b$, we conjecture that the effective field of each junction is changed with $b$, leading to the change of peak frequency [19]. Since the larger junction can reduce demagnetization energy more easily by forming complicated spin configurations, the effective field of larger junction becomes smaller. This is consistent with the main peak frequencies. Finally, let us pay our attention to the spectra at the center position D. According to Fig. 4 (c), the switching occurs abruptly, and the corresponding Fourier transform is broad spectra as shown in Fig. 5 (c). The Fourier spectra of position D shows no noticeable peak except broad structure in low frequency region (< 5 GHz), which is found in all position spectra. This is somewhat surprising results. Since the main peaks around 20 GHz correspond to the spin wave excitation, which is predicted by the macro spin model, it must play an important role in the switching processes. However, the switching process at the center position D has no specific frequency dependence. Therefore, we can conclude that the macro spin model is too simple to understand the real magnetization reversal process.

Figs. 6 (a)~(f) show the snapshots of specific times seen in Figs. 4 for $60 \times b$ nm$^2$. Figs. 6 (a) and (b) are the snapshots of $60 \times 20$ nm$^2$ at $t$ = 9.18 and 9.24 ns. As already discussed, the magnetization at positions A and B oscillates strongly, and at C it oscillates weakly. Furthermore, there is no oscillation at the center position D. Such behavior starts before 2 ns, and keeps steady oscillation till turn the current off. The motions of left and right sides are very asymmetry, so that the motion of center part is suppressed. Figs. 6 (c) and (d) of $60 \times 30$ nm$^2$ also show similar asymmetry along the long axis. However, there is asymmetry breaking along the short axis as shown in Fig. 6 (d). The asymmetry breaking is more significant for $60 \times 40$ nm$^2$ in Figs. 6 (e) and (f). At $t$ = 8.20 ns, the asymmetry is already broken and the center magnetization starts to move, and at $t$ = 9.24 ns, the center magnetization shows finite rotation. One possible reason of such asymmetry breaking is easy formation of complicated spin configuration due to the larger junction size.

However, the dependence of switching current density on the junction sizes is not simple as already shown in Fig. 2. The switching process requires asymmetry breaking at the center position, and it is strongly coupled with the detail spin configuration and dynamics which depends on the exchange stiffness and the junction size.

We investigate more details of the spin configurations during the switching process, but it is difficult to find simple relationship between the switching current and junction size. Based on our observations, however, we can make some conjectures. The magnetization dynamics along the long axis are asymmetric so that the STT effect on the center part is suppressed. Therefore, the $J_c$ dependence on the junction sizes is related with the formation of complicated spin configuration, which is determined by the exchange length, $l_{ex} \sim \pi\sqrt{A_{ex}/K_u} \sim \pi\sqrt{2A_{ex}/f\mu_0 M_s^2} \sim 10\sqrt{A_{ex}/f}$ nm, where $f$ is the correction factor of shape anisotropy (< 1) and $A_{ex}$ is in unit of $10^{-11}$ J/m. Therefore, the exchange length varies 10~20 nm, which depends on the exchange stiffness and the junction size in this study, and it is comparable to the junction dimension. This implies that the length scale of a few 10 nm can lead to noticeable variation of the shape anisotropy energy and it causes the limitation of spin wave vector $k$. This limitation is what we call weak quantization of spin wave vector $k$. Therefore, irregular variation of switching current for the exchange stiffness and the junction size is understandable.

It must be addressed that the asymmetry breaking can be more easily achieved in real experiments due to the non-uniform current density, which is not implemented in our simulations. Typical MTJ device has an order of ~100% tunneling magnetoresistance (TMR), it leads to the rapid variation of TMR from place to place. Since the electrodes are metallic, the potential across the insulating layer is equal, so that the local current density will vary from place to place. It depends on the relative orientation of magnetization between free and reference layers. Furthermore, imperfect junction shape introduced by the lithography processes also leads to the asymmetry breaking more easily. Therefore, the junction size dependence may be weaker than that expected.

We also investigate other cases in Figs. 2, however, the detail spin dynamics is too complicate to be explained with simple model. What we can claim is that the $J_c$ variation is much stronger than that estimated by macro spin model, and it requires more careful analysis.

## IV. CONCLUSIONS

We investigate the effect of junction size on the switching current density by employing micromagnetic simulations with STT. It is found that the dependence of the junction size is much stronger than that estimated by macro spin model. The variation of switching current densities can be explained by the formation of asymmetry breaking of spin configurations, which is determined by the exchange stiffness and shape anisotropy energies. Based on our micromagnetic simulations, we can conclude that the main reason of the large variation of the switching current densities is that the junction dimension is comparable to the exchange length of the system. Therefore, there is more chance to reduce the switching current density by optimization of the exchange stiffness and junction size.

**Acknowledgements**

This work was supported by the NRF funds (Grant Nos. 2010-0023798 and 2010-

**Figure Captions**

Fig. 1 Typical MTJ structures with synthetic antiferromagnetic fixed layer. The length of the long (short) axis of the ellipse is denoted by "*a*" ("*b*"). The labels "A"~"D" indicate the specific position of the junction along the long axis. "A" is the left end and "D" is the center of the ellipse.

Fig. 2 Switching current densities as a function of long axis length *a* for various short axis length *b* (= 20, 30, and 40 nm) with different exchange stiffness constant $A_{ex}$ of (a) 1.0, (b) 2.0, and (c) $3.0 \times 10^{-11}$ J/m.

Fig. 3 Switching current densities as a function of long axis length *a* for various exchange stiffness constant $A_{ex}$ ( = 1.0, 2.0, and $3.0 \times 10^{-11}$ J/m) with the length of short axes *b* of (a) 20, (b) 30, and (c) 40 nm.

Fig. 4 Time dependent normalized $M_x$ for each position A, B, C and D, and total $M_x$. The exchange stiffness constant is $A_{ex} = 1.0 \times 10^{-11}$ J/m and the junction dimensions are (a) $60 \times 20$, (b) $60 \times 30$, and (c) $60 \times 40$ nm$^2$ with the current density of $3.0 \times 10^{11}$ A/m$^2$.

Fig. 5 Fourier transform of the normalized $M_x$ of Fig. 4 for each position A, B, C, and D indicated in Fig. 1. The exchange stiffness constant is $A_{ex} = 1.0 \times 10^{-11}$ J/m and the junction dimensions are (a) $60 \times 20$, (b) $60 \times 30$, and (c) $60 \times 40$ nm$^2$ with the current density of $3.0 \times 10^{11}$ A/m$^2$.

Fig. 6 Snapshots of magnetization configurations at specific times of Fig. 4. (a, b) are for 60×20, (c, d) are for 60×30, and (e, f) are for 60×40 nm$^2$.

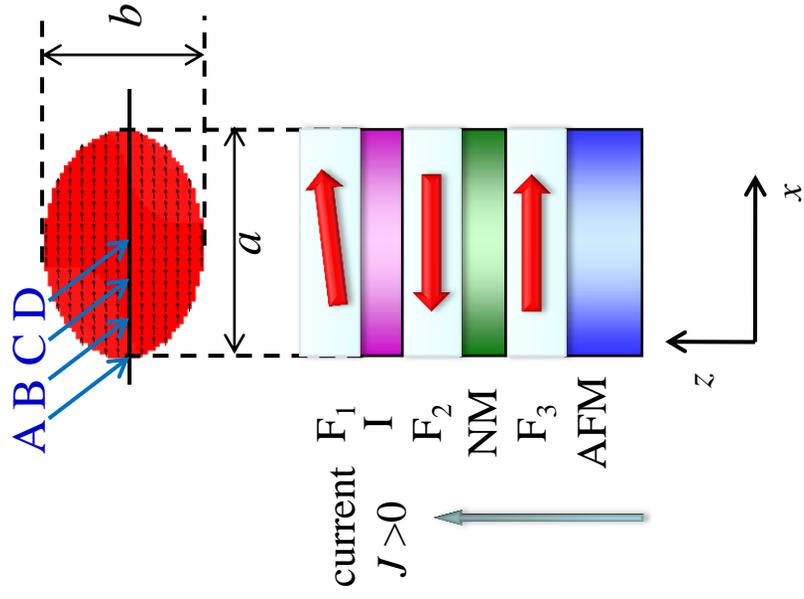

Fig. 1

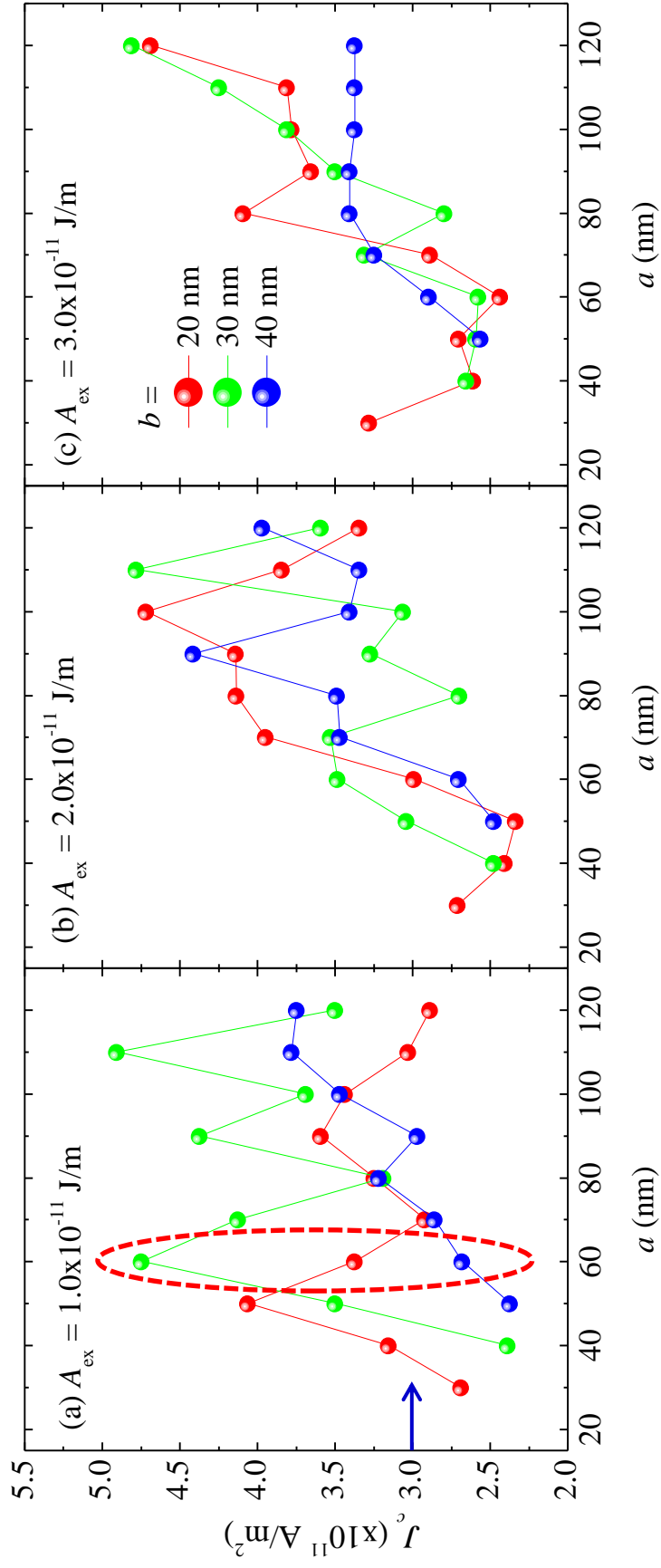

Fig. 2

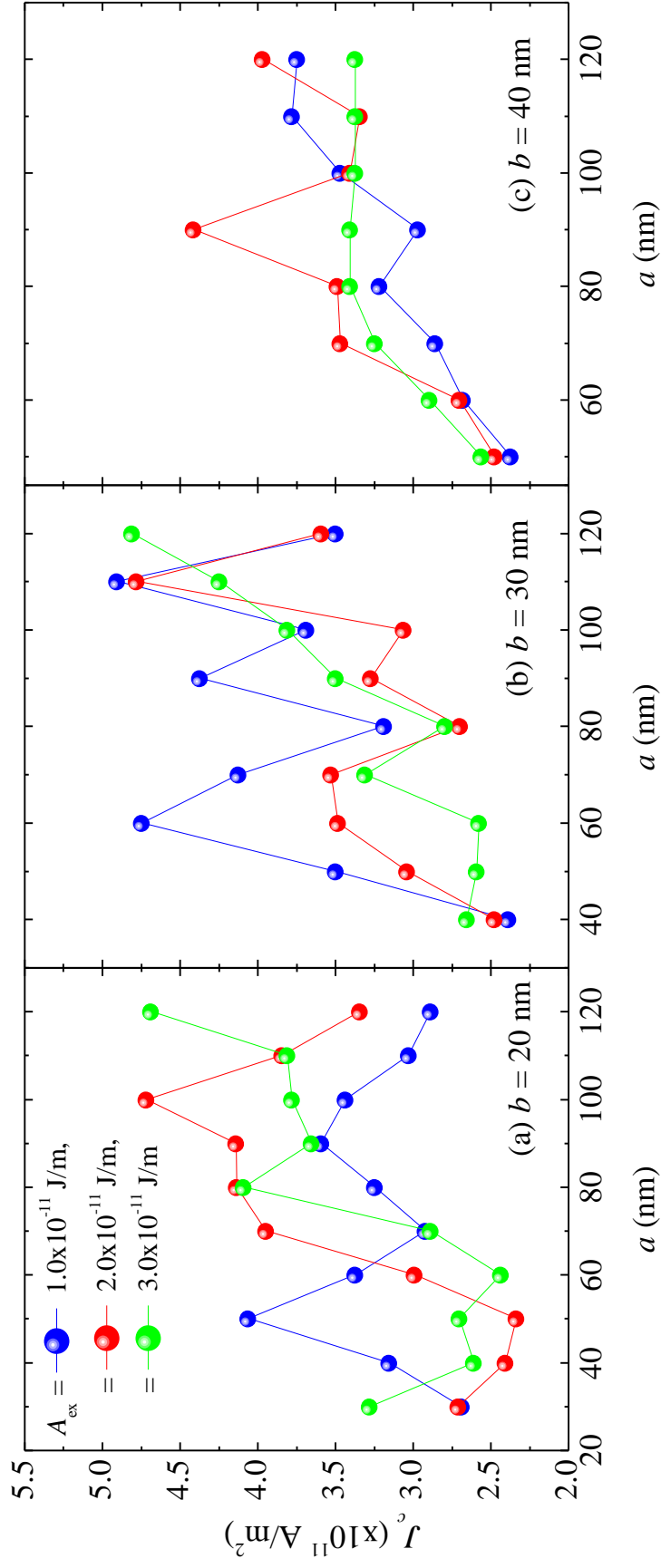

Fig. 3

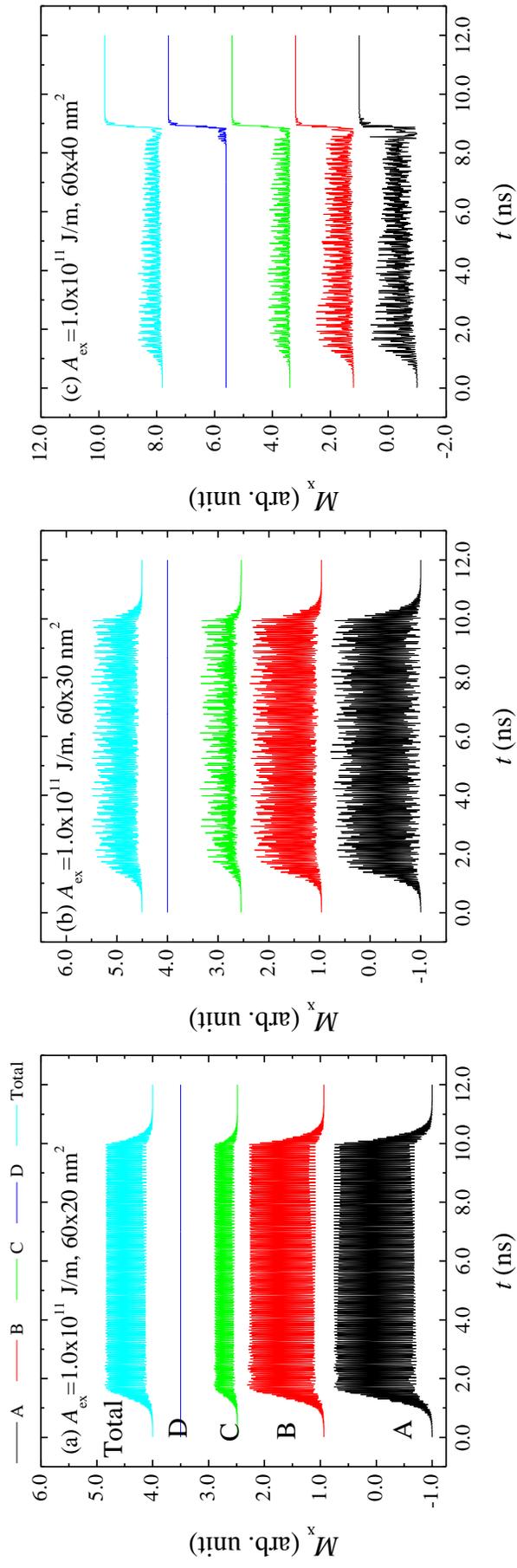

Fig. 4

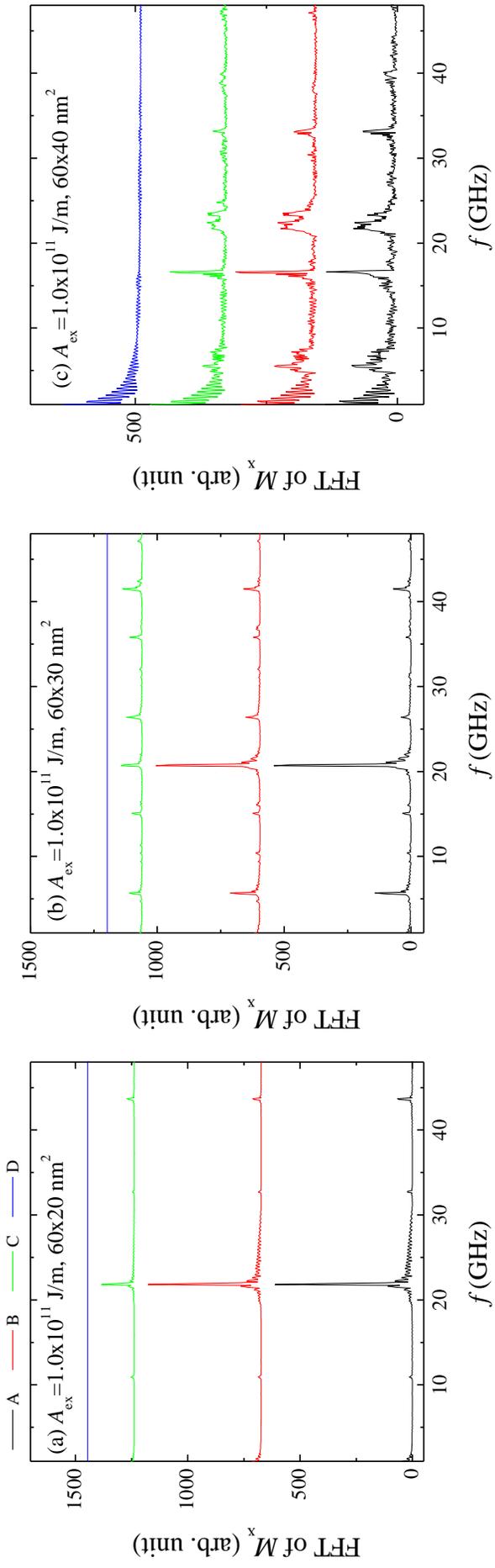

Fig. 5

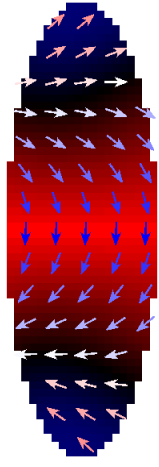
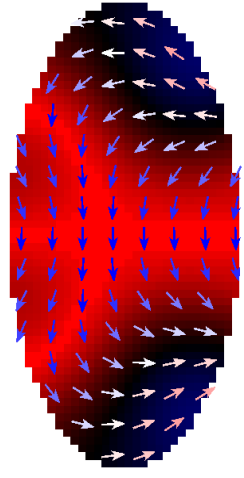
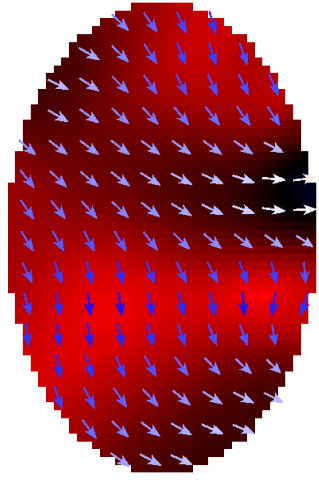
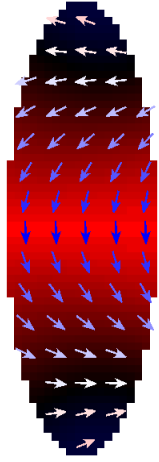
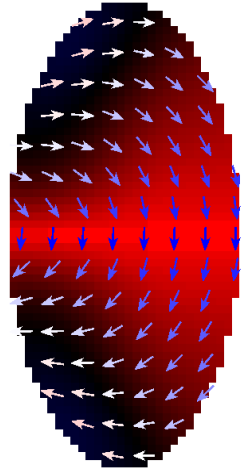
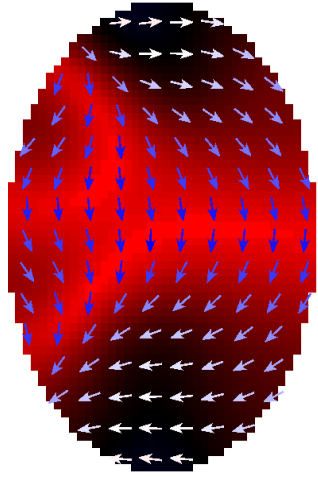

(a) $t = 9.18$ ns  (b) $t = 9.24$ ns   $60 \times 20$ nm$^2$

(c) $t = 9.13$ ns  (d) $t = 9.18$ ns   $60 \times 30$ nm$^2$

(e) $t = 8.20$ ns  (f) $t = 9.24$ ns   $60 \times 40$ nm$^2$

Fig. 6